\documentclass[aps]{JHEP3}

\usepackage{graphicx}
\usepackage{dcolumn}
\usepackage{amsfonts}
\usepackage{amsmath}

\newcommand{\pa}{\partial}
\newcommand{\td}{\tilde}

\newcommand{\beq}[1]{\begin{eqnarray}\label{#1}}
\newcommand{\eeq}{\end{eqnarray}}

\title{Strings in Noncommutative Spacetime}
\author{Xiao-Jun Wang\\
Interdisciplinary Center for Theoretical Study \\
University of Science and Technology of China \\
AnHui, HeFei 230026, China \\
E-mail:wangxj@ustc.edu.cn}

\abstract{Free bosonic strings in noncommutative spacetime are
investigated. The string spectrum is obtained in terms of
light-cone quantization. We construct two different models. In the
first model the critical dimension is still required to be 26
while only extreme high energy spectrum is modified by
noncommutative effect. In the second model, however, the critical
dimension is reduced to be less than 26 while low-energy
(massless) spectrum only contains degrees of freedom of our four
dimensional physics.}


\preprint{USTC-ICTS-05-05}
\begin{document}

\section{Introduction}

There is a long-standing belief that spacetime must change its
nature at the distance comparable to the Planck scale in any
quantum theories including gravity. A natural generalization of
quantum mechanism principle to quantum gravity is to introduce
uncertainty principle among spacetime coordinates which prevents
one from measuring positions to better accuracies than the Planck
length\cite{DeWitt62}: The momentum and energy required to perform
a measurement at these scales will modify the geometry of
spacetime. The simplest and naive realization of the above idea is
to postulate the commutation relations among spacetime coordinates
\beq{1.1}[x^i,x^j]=i\theta^{ij},
\eeq
with a constant parameter $\theta$ which is an antisymmetric
tensor. However, there are two questions which should be answered:
1) Could the commutation relations~(\ref{1.1}) be indeed
reproduced by a quantum theory including gravity? 2) How to study
gravity fluctuations of on the noncommutative backgrounds, i.e.,
to construct a noncommutative theory of quantum gravity?

To answer these questions, a consistent quantum theory including
gravity should be {\sl a priori} known. String theory, at least
perturbatively, may be such a candidate. Unfortunately,
perturbative string theory itself is failed to answer the first
question since the background has to be input in advance. The
current terminology ``noncommutative geometry in string theory''
often states as noncommutativity of world-volume coordinates of
D-branes\cite{LY96,CDS98,DH98,HV98} or equivalently of coordinates
of end-points of open strings\cite{CH98}. It induces various
noncommutative quantum field theories without gravity on
world-volume of D-brane. These theories can be viewed as a sort of
effective description on open strings ending on D-branes in a
background with large constant NS field along world-volume of
D-branes\cite{SW99}. This background does not directly change
anything of closed strings as well as of gravity at lower energy.
Instead, the amplitudes involving closed strings are changed
through coupling between closed string and open
string\cite{DT01,LM01,OO01}. These studies, therefore, can not
answer the second questions too.

Although there are many difficulties, we may ask an alternative
question: Is it possible to construct the perturbative string
theory on the background with commutation relations~(\ref{1.1})?
In this present paper we shall contribute a small step along this
direction, that is to study the light-cone quantization of bosonic
strings on the flat background with commutation
relations~(\ref{1.1}) and extract energy spectrum of string
oscillation. The progress is almost the same as standard progress
for strings on flat commutative background, except that the
additive noncommutativity of world-sheet scalar fields changes
commutators among raising and lower operators. Consequently the
energy spectrum of strings on the commutative background is
modified.

The directly generalization of Eq.~(\ref{1.1}) to the ``equal
time'' canonical commutation relations of world-sheet scalar
fields are as follows,
\beq{1.2}[X^i(\tau,\sigma),X^j(\tau,\sigma')]=\left\{
\begin{array}{ll}i\alpha'\theta^{ij}, &\hspace{0.6in}{\rm if}\quad
\sigma=\sigma', \\
0,&\hspace{0.6in}{\rm if}\quad \sigma\neq\sigma',
\end{array}\right.
\eeq
with $\tau,\;\sigma$ world-sheet coordinates and $\alpha'$ the
Regge slope. In equation~(\ref{1.2}) we rescale the antisymmetric
tensor $\theta$ so that it is dimensionless now. When $l_{nc}\sim
l_s$ with $l_{nc}$ the noncommutative length and
$l_s=\sqrt{\alpha'}$ the string length, however, there is an extra
complication: The commutation relations~(\ref{1.1}) implies an
uncertainty region order $l_{nc}$. The distance of any two points
inside this region can not be measured accurately. On the other
hand, $l_s\sim l_{nc}$ indicates that the whole of string is
included into such an uncertainty region, hence the positions of
any two points on strings are in general noncommutative,
\beq{1.3}[X^i(\sigma),X^j(\sigma')]=i\alpha'\theta^{ij} {\cal
F}(\sigma-\sigma').
\eeq
The commutation relations~(\ref{1.3}) yields large ambiguities to
build noncommtative models of strings. However, we may expect that
Eq.~(\ref{1.2}) is a sort of limit of (\ref{1.3}). It imposes a
strong constraint to function ${\cal F}(\sigma-\sigma')$.
Hereafter we shall call the models with commutation
relations~(\ref{1.2}) as ``rigorous noncommutative model'', and
those with relations~(\ref{1.3}) as ``fuzzy noncommutative
model''. These two kinds of models, as shown in our studies, yield
very different physics.

In section 2 of this paper, we study the light-cone quantization
of the rigorous noncommutative model. While a fuzzy noncommutative
model is investigated in section 3. It shares some basic
properties of any fuzzy models. Discussions and conclusions are
included in section 4.

\section{Rigorous noncommutative model}

The commutation relations~(\ref{1.2}) and (\ref{1.3}) do not
modify classical action of bosonic strings on $D$-dimensional flat
spacetime,
\beq{2.1} S=-\frac{1}{4\pi\alpha'}\int d\tau d\sigma\sqrt{-g}
g^{ab}\pa_aX^\mu\pa_bX^\mu,
\eeq
as well as classical equations of motion. Consequently the
conformal symmetry on world-sheet is kept at classical level at
least. Since quantum theories with noncommutativity in time may
have problem with unitarity\cite{GM00}, we will consider only
spatial noncommutativity, i.e., $\theta^{0i}=0,\;(i=1,2,...,D-1)$.
Then we can take light-cone gauge that is the same as one on
commutative background. The mode expansions of transverse
world-sheet scalar fields are also the same, i.e.,
\beq{2.2}X^i=x^i+\frac{p^i}{p^+}\tau+i\sqrt{\frac{\alpha'}{2}}
\sum_{n=-\infty\atop n\neq 0}^{\infty}
\left\{\frac{\alpha_n^i}{n}e^{-in(\tau+\sigma)}+
\frac{\td{\alpha}_n^i}{n}e^{-in(\tau-\sigma)}\right\},
\hspace{0.3in}\sigma\in [0,2\pi],
\eeq
for closed strings and
\beq{2.3}X^i=x^i+\frac{p^i}{p^+}\tau+i\sqrt{2\alpha'}
\sum_{n=-\infty\atop n\neq 0}^{\infty}\frac{\alpha_n^i}{n}
e^{-in\tau}\cos{n\sigma},\hspace{1in}\sigma\in [0,\pi]
\eeq
for open strings. It is not hard to check that the following
canonical commutators
\beq{2.4}[\alpha_n^i,\alpha_m^j]
&=&[\td{\alpha}_n^i,\td{\alpha}_m^j]=n\delta_{m+n,0}\left\{
\delta^{ij}+\frac{2i}{\pi^2}\theta^{ij}\lim_{L\to\infty}
\sum_{l=1}^L\frac{n}{n^2-(l-1/2)^2}\right\}, \nonumber \\
\mbox{} [x^i,p^j] &=& i\delta^{ij},
\hspace{1in}[x^i,x^j]=[p^i,p_j]=0
\eeq
yield the ``equal time'' canonical commutation relations (with
$\Pi^i$ canonical momenta)
\beq{2.5}[X^i(\sigma),\Pi^j(\sigma')]&=&
i\delta^{ij}\delta(\sigma-\sigma'), \nonumber \\
\mbox{}[X^i(\sigma),X^j(\sigma')]&=&\left\{
\begin{array}{ll}i\alpha'\theta^{ij}, &\hspace{0.6in}{\rm if}\quad
\sigma=\sigma', \\
0,&\hspace{0.6in}{\rm if}\quad \sigma\neq\sigma'
\end{array}\right.
\eeq
for both of closed and open strings. It is just that we wanted.
While commutation relations among $\Pi^i$ are divergent
\beq{2.6}[\Pi^i(\sigma),\Pi^j(\sigma')]\;\propto\;
iL\theta^{ij}\delta(\sigma-\sigma')+....
\eeq
In equation~(\ref{2.4}) we defined the infinite summation as a
limit since the summation
$$\sum_{l=1}^\infty\frac{n}{n^2-(l-1/2)^2}$$
is not well-defined for $n\to\infty$.

The ground states for oscillation of closed strings,
$|0,0;k\rangle$ can again be defined to be annihilated by the
lowering operators $\alpha_n^i,\;(n>0)$ and to be eigenstates of
the center-of-mass momenta. Furthermore, the general eigenstates
of Hamiltonian,
\beq{2.7}H=\frac{1}{2p^+}p^ip^i+\frac{1}{p^+\alpha'}
\sum_{n=1}^\infty(\alpha_{-n}^i\alpha_n^i
+\td{\alpha}_{-n}^i\td{\alpha}_n^i)+\frac{2-D}{12},
\eeq
are constructed by acting on $|0,0;k\rangle$ with the raising
operators,
\beq{2.8}|N,\td{N};k\rangle=\prod_{i=2}^{D-1}\prod_{n=1}^\infty
\frac{(\beta_{-n}^i)^{N_{in}}(\td{\beta}_{-n}^i)^{\td{N}_{in}}}
{(\lambda_{in}^{N_{in}+\td{N}_{in}}N_{in}!\td{N}_{in}!)^{1/2}}
|0,0;k\rangle,
\eeq
where
\beq{2.9}&&\left\{\begin{array}{l}\beta_{-n}=U\alpha_{-n}, \\
\beta_{n}^T=\alpha_{n}^TU^\dag,\end{array} \right. \hspace{0.8in}
\left\{\begin{array}{l}\td{\beta}_{-n}=U\td{\alpha}_{-n}, \\
\td{\beta}_{n}^T=\td{\alpha}_{n}^TU^\dag, \end{array} \right.
\hspace{0.6in}n>0, \nonumber \\
&&\lambda_{in}=n+\frac{2\nu_i}{\pi^2}\lim_{L\to\infty}
\sum_{l=1}^L\frac{n^2}{n^2-(l-1/2)^2}.
\eeq
with $U$ unitary matrix which diagonalizes anti-symmetrical matrix
$\theta$ and $\nu_i$ the eigenvalues of $i\theta$ (so that all of
$\nu_i$ are real).

Finally we obtain the energy spectrum,
\beq{2.11}E^2=k^ik^i+\frac{2}{\alpha'}\left\{
\sum_{n=1}^\infty\lambda_{in}
(N_{in}+\td{N}_{in})+\frac{2-D}{12}\right\},
\eeq
with level match condition
\beq{2.12}\sum_{n=1}^\infty\lambda_{in}
(N_{in}-\td{N}_{in})=0.
\eeq
Since
\beq{2.13}\sum_{l=1}^L\frac{n}{n^2-(l-1/2)^2}=\sum_{l=1}^{2n}
\frac{1}{2(l+L-n)-1},
\eeq
at the limit $L\to\infty$ energy spectra of finite $n$ modes do
not receive modifications from noncommutative effect. In other
words, the dimensions of spacetime is still required 26 and the
massless states\footnote{The noncommutativity of spacetime
violates the Lorentz symmetry, so that in general we can not
define what is ``mass'' of a particle. In this paper we will
usually refer ``energy spectrum'' instead of ``mass spectrum''. At
low energy, however, the Lorentz invariance is good symmetry so
that the massless states can be defined as usual.} are the same as
those in commutative case. For those modes with $n=q L\to\infty$,
$q\ll 1$, however, their energy spectra are modified due to
$\lambda_{in}\simeq n(1-4q\nu_{i}/\pi^2)$. It reflects a fact that
noncommutative effects can only be detected by extreme high energy
modes.

The above processes can be applied to open strings and the same
conclusions are obtained.

\section{Fuzzy noncommutative model}

The rigorous model studied in previous section has an obvious
inconsistency at extreme short distance: The
summation~(\ref{2.13}) is logarithmically divergent for $n/L\simeq
1,\; L\to\infty$. Since antisymmetric matrix $i\theta$ always has
negative eigenvalue, the modes with extreme short wave length (or
sufficient large $n$) have ill-defined energy $E^2\to -\infty$. It
supports the argument in Introduction that the commutation
relations~(\ref{1.2}) should be replaced by (\ref{1.3}) when
string scale is order to Plank scale. In this section we will
consider a simple fuzzy model in which $L$ in commutation
relations~(\ref{2.4}) is set to be finite integer. Hence the
rigorous model is a limit of this fuzzy model. Although this model
is very special, it shares some basic features of this type of
models.

For finite $L$, the summation~(\ref{2.13}) is also finite. So that
the noncommutativity yields a finite modification to whole string
spectrum except for mode with $n=0$. In particular, this
modification completely breaks the spatial rotation $SO(D-2)$
symmetry of massless states ($n=1$ modes) on original commutative
background. For general choice of antisymmetric matrix $\theta$,
there are no any massless states in this model. In order for this
model to describe the physics of our world, the eigenvalues of
$\theta$ must be degenerate. It is particular interest to consider
that the eigenvalues of $\theta$ is double degenerated, which
corresponds that $D\times D$ matrix $\theta$ can be reduced to a
diagonal block form as follows
\beq{3.1}\theta=\left(\begin{array}{cccc}
0&&& \\
&0&& \\
&&A& \\
&&&A \end{array}\right).
\eeq
with $A$ antisymmetric $(D/2-1)\times (D/2-1)$ matrix.
Consequently dimensions of spacetime must be even and coordinates
in this representation may be divided into three group: $D=(1+1)+
(D/2-1)+ (D/2-1)$. Coordinates in different groups are commutative
each other, while those in one of latter two groups are
noncommutative.

Let $\nu_{\rm min}<0$ be minimal eigenvalue of $i\theta$. In order
that there are massless states in $n=1$ modes, we require
\beq{3.2}1+\frac{2\nu_{\rm min}}{\pi^2}
\sum_{l=1}^L\frac{1}{1-(l-1/2)^2}=\frac{D-2}{24}.
\eeq
Therefore, the critical dimension in this model is modified to
$D\leq 24$ by noncommutative effect since $\nu_{\rm min}$ is order
one when string scale is order noncommutative scale. Furthermore,
to avoid to appearance of new states with negative energy square,
we should impose the condition $\lambda_{in}\geq 0$. Since maximal
value of summation~(\ref{2.13}) corresponds to $n=L$, one has
\beq{3.3}1+\frac{2\nu_{\rm min}}{\pi^2}
\sum_{l=1}^L\frac{L}{L^2-(l-1/2)^2}=1+\frac{\nu_{\rm min}}{\pi^2}
\sum_{l=1}^{2L}\frac{1}{l-1/2}>0.
\eeq
Equation~(\ref{3.3}) together with (\ref{3.2}) yield
\beq{3.4}\frac{8L}{4L^2-1}+\frac{D-26}{24}
\sum_{l=1}^{2L}\frac{1}{l-1/2}>0.
\eeq
The above equation impose a constraint between $L$ and critical
dimension $D$. In table 1 we list the critical dimensions admitted
by diverse values of $L$. We can see that only few choices to $L$
are allowed.

\begin{table}[hptb]
\setlength{\tabcolsep}{0.4in}
 \begin{tabular}{c|c}
 \hline
 $L$ & critical dimension $D$ \\ \hline
 $1$ & $6,\;8,...,24$ \\
 $2$ & $20,\;22,\;24$ \\
 $3$ & $22,\; 24$ \\
 $4,\;5$ &  $24$ \\
 $\geq 6$ & no consistent choice \\
\hline
 \end{tabular}
\centering
 \begin{minipage}{4in}
 \caption{The critical dimensions for fuzzy noncommutative string model
 admitted by diverse values of $L$.}
 \end{minipage}
\end{table}

Using commutators~(\ref{2.4}), the function ${\cal
F}(\sigma-\sigma')$ defined in commutation relations~(\ref{1.3})
can be obtained
\beq{3.5}{\cal F}(\sigma-\sigma')=\frac{2}{\pi}\sum_{l=1}^L
\left\{\frac{1}{(l-1/2)^2}-\frac{\pi}{l-1/2}
\sin{(l-\frac{1}{2})(\sigma-\sigma')}\right\}.
\eeq
Physically, we expect that accurate measure of position of a point
is much more difficult than accurate measure of distance between
two points even though these two points are both included in the
same uncertainly region. In other words, we expect that the fuzzy
model should satisfy
\beq{3.6}
q = \frac{1}{2\pi{\cal F}(0)}\int_{0}^{2\pi}d\sigma |{\cal
F}(\sigma)| \ll 1.
\eeq
For our special model, the above condition implies that $L$ should
be chosen as large as possible. Indeed, direct calculations show
that $q=0.42$ for $L=1$ and $q=0.12$ for $L=5$. So that $L=5$ is a
better choice which requires $D=24$.

No matter how many spacetime dimensions are admitted, at low
energy there are only four massless states generated by
oscillation of closed strings, and two massless states generated
by open strings. They can be interpreted as a graviton, a gauge
boson and two scalars in four dimensions. In other words, the
low-energy spectrum of fuzzy model may contains degrees of freedom
of our $(3+1)$-dimension physics only and exhibits a
$(3+1)$-dimension Lorentz invariance. This is a basic properties
of the fuzzy models. Those models, therefore, may be regards as an
alternative compactification of string theory.

\section{Summary and discussion}

To conclude, we investigated bosonic strings in a flat but
noncommutative spacetime. The string spectrum is obtained in terms
of light-cone quantization. The low-energy physics predicted by
rigorous model is almost the same as usual bosonic string in
commutative background. However, the rigorous model is
inconsistent when string scale is order noncommutative scale. We
further proposed a fuzzy model which may avoid the inconsistency
of the rigorous model at very short distance. The critical
dimensions and low-energy degrees of freedom in the fuzzy model,
however, are reduced by noncommutative effect. In particular, it
is possible to construct a model in which massless spectrum only
contains degrees of freedom of our four-dimensional world although
many extra dimensions are presented. This is typical nonlocal
effect: Physics at short distance produces observable effect at
long distance.

It should be asked what is physical meaning to consider a string
theory in noncommutative background: String theory itself is
nonlocal. While any perturbative quantum theory on noncommutative
spacetime must exhibit nonlocal effects too. Then is there problem
of double counting? If we believe that spacetime must change its
nature at very short distance by supplement of commutation
relations such as~(\ref{1.1}) and non-perturbation definition of
string theory (or M-theory) is a correct quantum theory to
describe our world with gravity, we should expect that the
spacetime with commutation relations~(\ref{1.1}) is a
non-perturbative vacuum of string theory. Hence it does make sense
to consider the perturbative excitations of strings on this
vacuum.

The further progresses may focus on CFT approach and
supersymmetric generalization of these noncommutative model.
Unfortunately, there is a technical trouble to construct CFT
approach: The operator equation derived from
\beq{4.1}0&=&\int [dX]\frac{\delta}{\delta X_\mu(z,\bar{z})}
\left[e^{-S}X^\nu(z',\bar{z}')\right]
\eeq
is highly nonlinear and hard to solve. It prevents us from
obtaining the operator product expansion. It is expected to
overcome this difficulty in future studies.

\acknowledgments{This work is partly supported by the NSF of
China, Grant No. 10305017, and through USTC ICTS by grants from
the Chinese Academy of Science and a grant from NSFC of China.}

\end{document}